\input ./style/arxiv-general.cfg
\documentclass[MSNbibl,nameyear,dvips]{arxstspdf}
\makeatletter
   \@ifpackageloaded{graphicx}{}{\usepackage{graphicx}}
\makeatother
\usepackage{flushend}
\usepackage{stfloats}
\volume{30}
\issue{2}
\pubyear{2015}
\firstpage{181}
\lastpage{183}
\doi{10.1214/15-STS519}% Updated by VTEXPTS2LaTeX.exe, 20.03.2015 14:28
\referstodoi{10.1214/14-STS487}
\docsubty{FLA}

\makeatletter
\newcommand{\bW}{\mathbf{W}}
\newcommand{\bh}{\mathbf{h}}
\newcommand{\bs}{\mathbf{s}}
\def\T{^{\mathrm{T}}}
\makeatother

\begin{document}
\begin{frontmatter}
\vspace*{12pt}
\title{Rejoinder}%\thanksref{T1}
% kai straipsnis turi susijusiu diskusiju ir rejoinder'iu
%\relateddois{T1}{Discussed in \relateddoi{d}{10.1214/00-STSXXX} ...;
%rejoinder at \relateddoi{r}{10.1214/00-STSXXXX}.}
\runtitle{Rejoinder}
%\pdftitle{}

\begin{aug}
% Corresponding author: Marc Genton - marc.genton@kaust.edu.sa% Updated by VTEXPTS2LaTeX.exe, 23.03.2015 12:45
%by VTEXPTS2LaTeX.exe, 20.03.2015 14:28
\author[A]{\fnms{Marc G.}~\snm{Genton}\corref{}\ead[label=e1]{marc.genton@kaust.edu.sa}}
\and
\author[B]{\fnms{William}~\snm{Kleiber}\ead[label=e2]{william.kleiber@colorado.edu}}
\runauthor{M.~G. Genton and W. Kleiber}
%\pdfauthor{}

\affiliation{King Abdullah University of Science and Technology and University of Colorado}

\address[A]{\fontsize{9.6}{11.6}{\selectfont Marc G. Genton is Professor, CEMSE Division, King Abdullah
University of Science and Technology,
Thuwal 23955-6900, Saudi Arabia \printead{e1}.}}
\address[B]{\fontsize{9.6}{11.6}{\selectfont William Kleiber is Assistant Professor, Department of
Applied Mathematics, University of Colorado,
Boulder, Colorado 80309-0526, USA \printead{e2}.}}
\end{aug}

% ABSTRACT
%\begin{abstract}
%\end{abstract}

% KEYWORDS
%\begin{keyword}
%\kwd{}
%\end{keyword}
\end{frontmatter}

We are grateful to the discussants for providing very valuable and
insightful comments. Next, we present our views on some of the comments
of the discussants and provide further discussion.

We thank Bevilacqua, Hering and Porcu (hereafter, BHP)
for bringing attention to the fundamental
problem of comparing multivariate models. Until now, almost all comparisons
between models have been relegated to empirical performance on specific
datasets,
whether it be performance on cokriging or particular scoring rules.
BHP introduce two theoretical approaches to comparing the flexibility of
multivariate frameworks: (A) assessing the size of allowable
co-located cross-correlation between processes, and (B)~a measure of
difference in allowed spatial (cross)-correlation at differing distances.

Regarding (A), BHP claim the bivariate Mat\'ern is less flexible than the
LMC in that there are nontrivial restrictions on the cross-correlation
coefficient for the bivariate Mat\'ern that are not present for the LMC.
We emphasize, however, that the bivariate Mat\'ern restrictions
are a \textit{characterizing} feature of the covariance class---no LMC
construction can allow for marginal and cross exact Mat\'ern behavior while
allowing for unrestricted choice of co-located cross-correlation. Rather,
it is a physical restriction on the covariance class, not a flexibility
restriction.

Most spatial modelers include a nugget effect in the statistical model,
$Y_i(\bs) = Z_i(\bs) + \varepsilon_i(\bs)$, where $Z_i(\bs)$ is
endowed with a
multivariate model, and $\varepsilon_i(\bs)$ is a white noise process
that is
uncorrelated with $Z_i(\bs)$. If $\varepsilon_i(\bs)$ is nontrivial with
variance $\tau_i^2$, then the
restrictions on the cross-correlation coefficient can be relaxed, the amount
depending on the magnitude of the nugget effect and sample size.
To see this, let $p=2$ and write the covariance matrix for two unit
variance processes
at $n$ locations\vspace*{1pt} $\{Z_1(\bs_1),\ldots,Z_1(\bs_n),Z_2(\bs_1),\ldots,Z_2(\bs_n)\}\T$ as\vadjust{\goodbreak}
$B \odot\Sigma$, where $\Sigma= \{C_{ij}(\bs_k,\bs_\ell)\}
_{i,j=1;k,\ell=1}^{2;n}$
and $B = (B_{ij})_{i,j=1}^2$ consists\vspace*{1pt} of four $n\times n$ block matrices.
For simplicity, assume $\tau_1 = \tau_2$, so that
$B_{12} = B_{21}$ are matrices populated by a constant $\rho_0$ and
$B_{11}=B_{22}$ are matrices of ones with diagonal $1 + \tau^2$. Note that
the case $\rho_0 = 1$ results in $B \odot\Sigma$ having the specified
multivariate dependence; if $\rho_0 > 1$, then the two processes can have
larger cross-correlation than allowed by the specified model.
The cases where $\rho_0 > 1$ are valid when $B_{12} = B_{11}^{1/2} K
B_{22}^{1/2}$, where $K$ is a contraction matrix (i.e., a matrix whose
singular values are bounded by unity); this follows from Proposition 1 of
\citet{KG13}. This is one feasible way to relax the restrictions
that are suggested by BHP's (A) criterion. We view BHP's (B) as an alternative
interesting route to comparing models, although it is still unclear what
improvements a modeler would expect to gain for various magnitudes of the
(B) criterion.

%%%%%%%%%%%%%%%%%%%%%%%%%%%%%%%%%%%%%%%%%%%%%%%%%%%%%%%%%%%%%%%%%%%%%%%%

Cressie et al.~focus on three main aspects: the importance of modeling
the nugget effect (which yields additional potential difficulties in
the multivariate context), the pseudo cross-variogram and alternative approaches
to building multivariate structures.

We focused our efforts on reviewing multivariate covariance functions,
not multivariate modeling, a~byproduct of which is that we left little
discussion
to the issue of modeling the nugget effect. For instance,
the underlying latent smooth process
$\bW$ of \citeauthor{C000} [(\citeyear{C000}), equation (4)] still
requires specification of the multivariate
structure, regardless of whether a nugget effect will or will not
ultimately be
included. Nonetheless, these authors bring up an important point in that,
especially for multivariate processes, some variables may be measured by
the same instrument, in which case it may be expected that measurement
errors are correlated across variables at individual locations.
Disentangling microscale variability of the process from measurement error
is indeed a difficult prospect; Sang, Jun and Huang (\citeyear{SJH11}) used a full-scale
approximation for multivariate processes that explicitly breaks up large
scale, small scale and measurement error variability.\vadjust{\goodbreak}

Cressie et al.~champion a traditional geostatistical approach to estimation,
using a weighted least squares distance from empirical pseudo
cross-variograms to
estimate parameters of a parametric class of cross-covariance functions.
Although this is a feasible route to estimation,
interpretation of the pseudo cross-va\-riogram remains unclear,
unless the process is stationary and the variables are standardized, so
that the pseudo cross-variogram can be rewritten $\operatorname{Var}\{Z_1(\bs+
\bh)/\sigma_1 -
Z_2(\bs)/\sigma_2\}$, where $\sigma_i$ is the standard deviation of the
$i$th process. This seems a particularly important consideration if
the scales of the two processes differ by orders of magnitude.
A more modern approach to estimation (albeit one that requires more modeling
assumptions) is to adopt a maximum likelihood or Bayesian framework.
Even if the processes are not strictly Gaussian, say,
these other approaches can still be used for estimation if the deviation
from Gaussianity is not too great.

We agree with Cressie et al.~that the conditional approach to estimation
has utility in certain situations when there is clear directional
dependence between variables. However, if many (say, greater than four)
processes are considered simultaneously, we echo Cressie et al.'s caution:
``when more variables are involved, the order may not always be obvious but,
if the goal is to construct valid covariance and cross-covariance functions,
the different orderings can be viewed as enlarging the space of valid models.''
On the other hand, the factor process approach seems promising. Given
the difficulties basis decomposition approaches experience in the univariate
setting (\cite{Finetal09,Ste14}), we expect similar issues to
arise in the multivariate setting. Thus, we suggest a multiresolution
approach, decomposing differing scales of support into various
resolutions of
basis functions (\citeauthor{Nycetal14}, \citeyear{ny02,Nycetal14}).

Finally, Cressie et al.~bring up some important issues in validating and
comparing statistical models that we view as sensible guidelines for future
authors working in this field. Indeed, direct likelihood comparisons
may be
muddled by the varying numbers of parameters between models, and BHP
have introduced some tools apart from the usual information criteria
to compare multivariate models.

%%%%%%%%%%%%%%%%%%%%%%%%%%%%%%%%%%%%%%%%%%%%%%%%%%%%%%%%%%%%%%%%%%%%%%%%

We are pleased that Simpson, Lindgren and Rue (hereafter, SLR) decided
to expand on
our very brief mention of the spectral representation of the
cross-covariance matrix function,
which for simplicity was restricted to the symmetric case in our
review. Although we did address the topic of
asymmetric cross-covariance functions in Section~5.1, SLR now provided
information about
the spectral representation in the asymmetric setting and further
discussed the spectral representation
of multivariate Gaussian random fields themselves. SLR also argued that
this path leads naturally
to non-Gaussian or nonstationary multivariate random fields, and they
further made the link with
physics-constrained cross-covariance models.

Although SLR advocated the SPDE approach as a computationally efficient
reformulation of univariate
and multivariate Gaussian random fields, there are also limitations
that can only be exacerbated in
the multivariate setting. For example, the smoothness parameter of the
Mat\'ern covariance function
is restricted to certain values. Moreover, this methodology requires a
strong background in numerical analysis techniques
which is not common among statisticians.
Thus, we view the multivariate kernel convolution approach as a compromise
where physics-based information can be incorporated, without
confronting the
numerous difficulties in implementing a SPDE framework.
Nevertheless, the route of SPDEs offers many interesting and
challenging avenues
for future research.

SLR end their discussion with an important point about the inflation of
the number of parameters in
cross-covariance models as the number of variables increases. This is
indeed a challenging issue
when applying likelihood methods and various authors have resorted to
composite likelihood approaches,
although more investigations in this area are warranted. Finally, the
problem of parameter identifiability under infill asymptotics in the
multivariate setting is mentioned and we see no reason for this effect
to disappear in this context.

%%%%%%%%%%%%%%%%%%%%%%%%%%%%%%%%%%%%%%%%%%%%%%%%%%%%%%%%%%%%%%%%%%%%%%%%

The discussion of Zhang and Cai (hereafter, ZC) centers on trying to
understand why and when cokriging does not always
outperform kriging. This is a very relevant topic given the numerical
results provided in our two data analyses.
To address this topic, ZC start by deriving sufficient conditions for
the equivalence of Gaussian measures for
a bivariate Mat\'ern cross-covariance model with common length scale
and smoothness parameters, but allowing
for different marginal variances. Using ZC's notation, an example of
two bivariate Gaussian measures that satisfy their
sufficient conditions is given by $\nu=1$, $\sigma_{11,1}=\sigma
_{22,1}=1$, $\sigma_{11,2}=\sigma_{22,2}=2$, $\sigma_{12,1}=1/2$,
$\sigma_{12,2}=1$, $\alpha_1=2$, and $\alpha_2=1$.

The example provided by ZC where cokriging is equivalent to kriging
falls into the category of autokrigeability
(Wackernagel, \citeyear{W03}, page 149). A~variable is autokrigeable with respect
to a set of variables if the kriging of this variable
is equivalent to the cokriging. A trivial case is when all variables
are uncorrelated. Another case, as illustrated by ZC,
is when the cross-covariance function is separable (also called
intrinsically correlated in geostatistics). Generalizations of
the concept of autokrigeability to various simplifications of large
cokriging systems by means of screen effects were investigated by
Subramanyam and Pandalai (\citeyear{SP08}); see also Furrer and Genton (\citeyear{FG11}) for
related methods to handle highly multivariate spatial data.

ZC conclude by investigating a situation where the auxiliary variable
is observed at more locations than the predicted variable,
leading to the cokriging predictor being more efficient than the
kriging predictor as a function of the co-located correlation
coefficient, $r$, in a separable exponential cross-covariance function
setting. Interestingly, their formula (9) shows that
the mean squared prediction error of the cokriging predictor can be
reduced to at most $1/2$ of that of the kriging predictor in this
particular case. We conjecture that the factor $1/2$ could be further
reduced by either adding more observation points in $O$ or by adding
more variables $Y_3(\bs),\ldots,Y_p(\bs)$ in a similar
cross-covariance function framework.

%%%%%%%%%%%%%%%%%%%%%%%%%%%%%%%%%%%%%%%%%%%%%%%%%%%%%%%%%%%%%%%%%%%%%%%%

Finally, one last issue that remains is the availability of statistical software
for implementing, simulating and estimating multivariate models.
Indeed, the
choice of a particular model to use in any application requires fair expertise
in spatial statistics, and we expect these models to become more mainstream
as well-documented software becomes more available. There are some promising
packages available currently that end users should be aware of, all of
which are available in \texttt{R}. Whereas \texttt{RandomFields}
contains a large
number of multivariate models and can perform high resolution simulation
of these (\citeauthor{Schetal14}, \citeyear{Schetal14}),
\texttt{spBayes} is geared toward Bayesian analyses of hierarchical
multivariate models (\cite{FinBanGel15}),
and \texttt{gstat} contains some variogram-based
estimation routines for the linear model of coregionalization (\cite{Peb04}).

%\begin{appendix}
%\section{}
%\end{appendix}

%\section*{Acknowledgments}

%\begin{supplement}[id=suppA]
%\slink[doi]{10.1214/00-STSXXXXSUPP} %[doi,text={...}] - jei reikia
%suskaldyti doi
%\sfilename{stsXXXX\_supp.pdf}
%\end{supplement}

% imsref loaded by linak, 2015-03-20 14:38:38
% imsref loaded by linak, 2015-03-20 14:42:51
% imsref loaded by linak, 2015-04-20 11:54:42
% imsref loaded by akundreckaite, 2015-05-11 12:28:31

\end{document}